\documentclass[journal=jpclcd,manuscript=letter]{achemso}
\usepackage[T1]{fontenc}
\usepackage[utf8]{inputenc}
\usepackage{amsmath}
\usepackage{graphicx}
\usepackage{placeins}
\PassOptionsToPackage{version=3}{mhchem}
\usepackage{mhchem}
\usepackage{xcolor}
\usepackage{commath}
\usepackage{subcaption}
\usepackage{multirow}
\usepackage{booktabs}
\usepackage{silence}
\usepackage{array}
\usepackage{braket}
\usepackage{placeins}
\usepackage{lineno}

\makeatletter

\author{Jiachen Li}
\affiliation{Department of Chemistry, Duke University, Durham, NC 27708, USA}
\author{Weitao Yang}
\affiliation{Department of Chemistry, Duke University, Durham, NC 27708, USA}
\email{weitao.yang@duke.edu}

\title{Renormalized Singles with Correlation in $GW$ Green's Function Theory for Accurate Quasiparticle Energies}

\makeatother

\begin{document}

\begin{abstract}
We apply the renormalized singles with correlation (RSc) Green's function in the $GW$ approximation to calculate accurate quasiparticle (QP) energies and orbitals. 
The RSc Green's function includes all orders of singles contributions from the associated density functional approximation (DFA) and treats higher order contributions in a perturbative manner.
The $G_{\text{RSc}}W_{\text{RSc}}$ method uses the RSc Green's function as the new starting point and calculates the screened interaction with the RSc Green's function. 
The $G_{\text{RSc}}W_0$ method fixes the screened interaction at the DFA level. 
For the calculations of ionization potentials in the GW100 set, 
$G_{\text{RSc}}W_{\text{RSc}}$ significantly reduces the dependence on the starting point of DFAs used and provides accurate results with the mean absolute error (MAE) of $0.34$ \,{eV} comparable to ev$GW$.
For the calculations of core-level binding energies in the CORE65 set,
$G_{\text{RSc}}W_{\text{RSc}}$ slightly overestimates the results because of underscreening, 
but $G_{\text{RSc}}W_0$ with GGA functionals provides the optimal accuracy of $0.40$ \,{eV} MAE comparable to ev$GW_0$.
We also show that $G_{\text{RSc}}W_{\text{RSc}}$ predicts accurate dipole moments of small molecules. 
These two methods, $G_{\text{RSc}}W_{\text{RSc}}$ and $G_{\text{RSc}}W_0$, are computationally much favorable than any flavor of self-consistent $GW$ methods.
Thus, the RSc approach is promising for making $GW$ and other Green's function methods efficient and robust.
\end{abstract}

\begin{tocentry}
\includegraphics[width=1\textwidth]{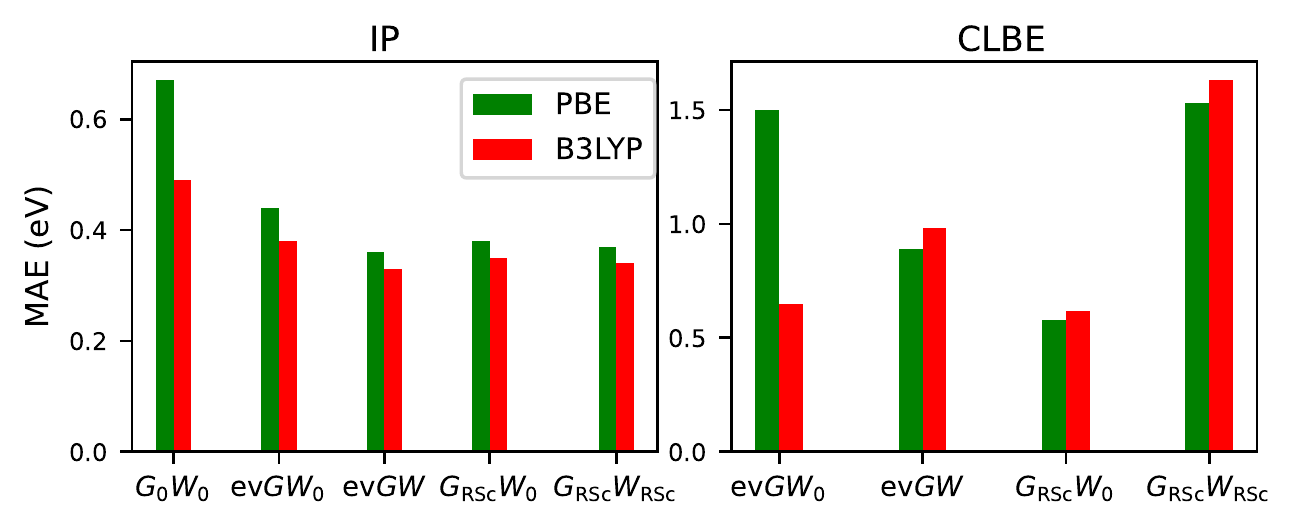}
\end{tocentry}

Quasiparticle (QP) energy is one of the most important quantities for studying molecules and materials.
It describes the charged excitation associated with the electron addition or removal,
which can be directly measured by the photoemission and inverse photoemission spectroscopy\cite{plummerAngleResolvedPhotoemissionTool1982,barrModernESCAPrinciples2020}.
For example,
the QP energies of valence states describe the ionization potentials (IPs) and the QP energies of core states describe the core-level binding energies (CLBEs).
These quantities play a critical role in characterizing the chemical environment for a broad range of systems\cite{beardCelluloseNitrateBinding1990,hofftElectronicStructureSurface2006} and guiding new applications in chemistry, biochemistry and material science\cite{fukagawaRoleIonizationPotential2007,galloDFTStudiesFunctionalized2007}. 
In past decades, 
much effort has been devoted to develop accurate and efficient theoretical methods to predict QP energies. 
The main workhorse in the electronic structure theory, 
Kohn-Sham density functional theory\cite{hohenbergInhomogeneousElectronGas1964,parrDensityFunctionalTheoryAtoms1989,kohnSelfConsistentEquationsIncluding1965} (KS-DFT), has been widely used in the quantum chemistry community to calculate total energies, electron densities and structures of both molecules and solids. 
Although the HOMO and LUMO energy in KS-DFT can be interpreted as the negative of the first IP and the first electron affinity (EA)\cite{cohenFractionalChargePerspective2008}, 
the errors of using KS eigenvalues to predict IPs and EAs are around several electronvolts\cite{vansettenGWMethodQuantumChemistry2013}.
The errors of KS-DFT with conventional density functional approximations (DFAs) for predicting CLBEs can even exceed $15$ \,{eV}\cite{golzeCoreLevelBindingEnergies2018}. 

In past decades, 
the $GW$ formalism has gained increasing attention.
The $GW$ approximation, derived from the Hedin's equations in the Green's function many-body perturbation theory\cite{hedinNewMethodCalculating1965,martinInteractingElectrons2016},
is a state-of-art formalism to compute QP energies and study spectral properties of charged excitations,
where $G$ is the one-body Green's function with poles at charged excitation energies and $W$ is the dynamical screened interaction.
The screened interaction that is generally weaker than the bare Coulomb interaction provides a better description for real systems because the effective interaction between two charges are reduced by the rearrangements of all other charges\cite{martinInteractingElectrons2016,golzeGWCompendiumPractical2019}. 
The $GW$ methods have been successfully implemented in modern quantum chemistry packages to study both valence and core properties of different systems ranging from molecules to solids\cite{renAllelectronPeriodic0W2021,maromBenchmarkGWMethods2012,knightAccurateIonizationPotentials2016,carusoSelfconsistentGWAllelectron2013,zhuAllElectronGaussianBasedG0W02021,zhuInitioFullCell2021}. 
The success of $GW$ methods can be attributed to the clear physical interpretation for QP energies, 
the favorable computational scaling with respect to the size of systems and the proper description of the interaction in real systems. \\

In practice, $GW$ methods are implemented at different levels of the self-consistency. 
The commonly used one-shot $G_0W_0$ method has been shown to significantly outperform KS-DFT for predicting valence and core QP energies\cite{vansettenGWMethodQuantumChemistry2013,golzeCoreLevelBindingEnergies2018}. 
However, $G_0W_0$ has a strong starting point dependence on the choice of the DFA because of its perturbative nature.
The difference originated from using different DFAs can exceed $1.0$ \,{eV} for IPs and EAs of molecules\cite{maromBenchmarkGWMethods2012,keAllelectronGWMethods2011}
and can be even larger than $5.0$ \,{eV} for band gaps of solids\cite{fuchsQuasiparticleBandStructure2007}. \\

The issue in $G_0W_0$ with conventional DFAs the systematic delocalization error\cite{mori-sanchezDiscontinuousNatureExchangeCorrelation2009,cohenInsightsCurrentLimitations2008}.
In comparison, Hartree-Fock (HF) approximation\cite{slaterNoteHartreeMethod1930,szaboModernQuantumChemistry2012} exhibits the localization error, 
opposite to the delocalization error. 
The delocalization error is manifested for small molecules as the convex deviation from the Perdew–Parr–Levy–Balduz (PPLB) condition\cite{perdewDensityFunctionalTheoryFractional1982,perdewExchangeCorrelationOpen2007,yangDegenerateGroundStates2000,cohenInsightsCurrentLimitations2008} that the total energy of a system as a function of the electron number should be piecewise linear between energies at integer points. 
However, for bulk systems and very large finite systems, the PPLB condition is satisfied for any DFA and the delocalization error is manifested as the incorrect slope of the straight lines of the total energy between integers\cite{mori-sanchezDiscontinuousNatureExchangeCorrelation2009,cohenInsightsCurrentLimitations2008}. \\

It has been shown that $G_0W_0$ starting based on a PBE-based hybrid functional with $40\%$ the HF exchange has the minimal deviation from the
straight-line error\cite{dauthPiecewiseLinearityGW2016} (DSLE) and is optimal for valence state calculations\cite{dauthPiecewiseLinearityGW2016,carusoBenchmarkGWApproaches2016}. 
$G_0W_0$ based on range-separated functionals was also shown to provide improved accuracy for computing valence spectrum\cite{korzdorferAssessmentPerformanceTuned2012,refaely-abramsonQuasiparticleSpectraNonempirical2012}.
For core state calculations, 
it has been shown that $G_0W_0$ with traditional DFAs gives multiple solutions to the QP equation of core states,
because the spectral weight of the desired core state is erroneously transferred to the satellites\cite{golzeCoreLevelBindingEnergies2018,golzeAccurateAbsoluteRelative2020,li2022benchmark}. 
The proper redistribution of the spectral weight can be achieved by tuning the fraction of the HF exchange in the starting point. 
Similar to the idea of DSLE, recent works show that the PBE-based hybrid function with $45\%$ the HF exchange is the optimal starting point for predicting CLBEs of molecules and semiconductors\cite{golzeAccurateAbsoluteRelative2020,zhuAllElectronGaussianBasedG0W02021,li2022benchmark}. 
One path to improve the accuracy and eliminate the starting point dependence is introducing the self-consistency into $GW$ calculations. 
For valence state calculations, 
fully self-consistent $GW$ (sc$GW$) \cite{rostgaardFullySelfconsistentGW2010,wangFullySelfconsistentSolution2015,carusoSelfconsistentGWAllelectron2013,carusoUnifiedDescriptionGround2012,holmFullySelfconsistentMathrmGW1998,stanLevelsSelfconsistencyGW2009,kovalFullySelfconsistentGW2014,gattiUnderstandingCorrelationsVanadium2007},
quasiparticle-self-consistent $GW$\cite{vanschilfgaardeQuasiparticleSelfConsistentGW2006,shishkinAccurateQuasiparticleSpectra2007,shishkinSelfconsistentGWCalculations2007,kaplanQuasiParticleSelfConsistentGW2016,brunevalEffectSelfconsistencyQuasiparticles2006,kotaniQuasiparticleSelfconsistentGW2007} (qs$GW$)
and eigenvalue-self-consistent $GW$\cite{kaplanQuasiParticleSelfConsistentGW2016} (ev$GW$) methods have been shown to significantly outperform $G_0W_0$ with the minimal starting point dependence.
For core states, ev$GW$ and qs$GW$ slightly overestimate CLBEs of molecular systems because of underscreening\cite{vansettenAssessingGWApproaches2018,li2022benchmark,golzeAccurateAbsoluteRelative2020}. 
The underscreening error can be largely compensated in the ev$GW_0$ method by fixing the screened interaction at the DFT level\cite{golzeAccurateAbsoluteRelative2020,li2022benchmark} because of the error compensation from the underestimated fundamental gap at the DFT level.
However, 
these partial self-consistent $GW$ methods also bring additional computational cost.
Techniques including the contour deformation approach\cite{golzeCoreLevelBindingEnergies2018,zhuAllElectronGaussianBasedG0W02021} and the analytical continuation approach\cite{ducheminRobustAnalyticContinuationApproach2020}
have been developed to reduce the computational cost of $GW$ calculations.
Recently the cubic scaling implementations\cite{wilhelmGWCalculationsThousands2018,ducheminCubicScalingAllElectronGW2021} of $GW$ calculations have also gained increasing attentions. 
Our previous work has shown that a simplified version of ev$GW_0$, 
the Hedin shift scheme\cite{li2022benchmark}, 
predicts accurate CLBEs with the comparable accuracy to ev$GW_0$. 
The $G_{\Delta \text{H}}W_0$ method shifts all eigenvalues by a global Hedin shift determined by the difference between the self-energy and the exchange-correlation KS potential of the desired state\cite{li2022benchmark}. 
Another path to compensate the underscreening error is to generalize the two-point screened interaction in $GW$ to a four-point effective interaction, 
which is the T-matrix approximation\cite{martinInteractingElectrons2016,zhangAccurateQuasiparticleSpectra2017,liRenormalizedSinglesGreen2021,romanielloGWApproximationCombining2012}. 
The T-matrix methods have been shown to predict accurate IPs and CLBEs of molecular systems\cite{zhangAccurateQuasiparticleSpectra2017,liRenormalizedSinglesGreen2021}. \\

Recently the renormalized singles (RS) Green's function\cite{jinRenormalizedSinglesGreen2019} was developed in our group to improve the accuracy and eliminate the starting point dependence in $G_0W_0$ with affordable computational cost. 
The RS Green's function that uses the form of the HF self-energy to capture all singles contributions is a good starting point for $GW$ calculations\cite{jinRenormalizedSinglesGreen2019,li2022benchmark}.
The RS Green's function has been applied in two methods:
$G_{\text{RS}}W_0$\cite{jinRenormalizedSinglesGreen2019} that uses the RS Green's function as the starting point and formulates the screened interaction with the KS Green's function and $G_{\text{RS}}W_{\text{RS}}$\cite{li2022benchmark} that uses the RS Green's function as the starting point and formulates the screened interaction with the RS Green's function.
$G_{\text{RS}}W_0$ outperforms $G_0W_0$ for predicting IPs with a small starting point dependence\cite{jinRenormalizedSinglesGreen2019} but fails to provide the unique solution to the QP equation for core states\cite{li2022benchmark}.
$G_{\text{RS}}W_{\text{RS}}$ provides consistent improvements over $G_0W_0$\cite{li2022benchmark} and has been combined with Bethe-Salpeter equation to compute accurate excitation energies (unpublished results).
The concept of RS has also been successfully applied in the multireference DFT approach\cite{liMultireferenceDensityFunctional2022} to describe the static correlation in strongly correlated systems.
The RS Green's function shares similar thinking as the renormalized single-excitation (rSE) in the random phase approximation (RPA) calculation for accurate correlation energies\cite{renRandomPhaseApproximationElectron2011, renRenormalizedSecondorderPerturbation2013, paierAssessmentCorrelationEnergies2012}. \\

In the present work, 
we go beyond the RS method to include the electron correlation in Green's function for self-energy, 
aiming to further improve the accuracy and eliminate the starting point dependence of $G_{\text{RS}}W_0$ and $G_{\text{RS}}W_{\text{RS}}$. 
Although the RS Hamiltonian that constructs the HF Hamiltonian with KS orbitals completely captures all singles contributions,
correlation contributions are not considered. 
The remaining starting point dependence in $G_{\text{RS}}W_0$ and $G_{\text{RS}}W_{\text{RS}}$ indicates that the correction for including higher order contributions is needed. 
As shown in Ref\citenum{li2022benchmark}, 
$G_{\text{RS}}W_{\text{RS}}$ overestimates CLBEs because the underscreening error.
The fundamental gap obtained from RS eigenvalues is overestimated and leads to an underscreened interaction, 
which is similar to $G_0W_0@$HF and ev$GW$\cite{maromBenchmarkGWMethods2012,li2022benchmark}. 
Therefore, the correction to the RS Hamiltonian needs to properly reduce the fundamental gap. 
We introduce the renormalized singles with correlation (RSc) Green's function. 
The RSc Green's function is obtained in a two-shot manner.
First, a $G_{\text{RS}}W_{\text{RS}}$ calculation is performed to obtain the correlation self-energy. 
Then the correlation self-energy is added to the RS Hamiltonian as a perturbative correction.
The RSc Green's function is constructed with the eigenvalues from the RSc Hamiltonian and is used in the second-shot calculation.
Similar to $G_{\text{RS}}W_0$ and $G_{\text{RS}}W_{\text{RS}}$,
the RSc Green's function can be applied in two methods: $G_{\text{RSc}}W_0$ and $G_{\text{RSc}}W_{\text{RSc}}$.
In the RSc Hamiltonian, 
exchange contributions are included completely and correlation contributions are treated in a perturbative manner. 
Compared with the self-consistent $GW$ methods that solve the Hedin's equations iteratively,
the RSc scheme provides a simple path to eliminate the starting point dependence in $GW$ formalism. 
The $G_{\text{RSc}}W_0$ and the $G_{\text{RSc}}W_{\text{RSc}}$ methods are analogous to ev$GW_0$ and ev$GW$ but are computationally much more favorable, 
without the need for self consistency. 
The RSc Green's function can also be applied in other Green's function methods such as T-matrix methods.\\

As shown in our previous work\cite{jinRenormalizedSinglesGreen2019,li2022benchmark,liRenormalizedSinglesGreen2021} and the Section~1 in the Supporting Information,
the RS Green's function is 
\begin{equation}
    [G_{\text{RS}}]_{pq} (\omega) = \delta_{pq}
    \frac{1}{\omega - \epsilon_{p}^{\text{RS}} +
    i\eta \text{sgn} (\epsilon_{p}^{\text{RS}} - \mu)} \text{,}
\end{equation}
where $\eta$ is an infinitesimal positive number and $\mu$ is the chemical potential. 
We use $i$, $j$ for occupied orbitals, $a$, $b$ for virtual orbitals, $p$, $q$ for general orbitals.\\

In the $G_{\text{RS}}W_{\text{RS}}$ method, 
$G_{\text{RS}}$ is used as the starting point and in the calculation of the screened interaction.
The exchange self-energy is the same as $G_0W_0$ because KS orbitals are used instead of RS orbitals for simplicity\cite{li2022benchmark}.
The correlation self-energy has a similar form as $G_0W_0$
\begin{equation}
        \Sigma^{c,G_{\text{RS}}W_{\text{RS}}}_{pq} (\omega) = 
        \sum_{m} \sum_{r} \frac{ \langle \psi^0_p \psi^0_{r} | \rho^{\text{RS}}_m \rangle \langle \psi^0_q \psi^0_{r} | \rho^{\text{RS}}_m \rangle }{\omega - \epsilon^{\text{RS}}_{r}
        - (\Omega^{\text{RS}}_m - i\eta) \text{sgn} (\epsilon^{\text{RS}}_{r} - \mu)} \text{, }
    \label{eq:grswrs_se}
\end{equation}
where $\rho^{\text{RS}}_m$ and $\Omega^{\text{RS}}_m$ are the transition densities
and the corresponding excitation energies from RPA calculated with the RS Green's function, 
$m$ is the index of the RPA excitation.\\

To further improve the accuracy of $G_{\text{RS}}W_{\text{RS}}$, 
the two-shot RSc scheme is introduced. 
In the first shot, 
the $G_{\text{RS}}W_{\text{RS}}$ calculation is performed to obtained the correlation self-energy defined in Eq.\ref{eq:grswrs_se}.
Then in the second shot,
$\Sigma^{c,G_{\text{RS}}W_{\text{RS}}}$ is used as the correction to the RS Hamiltonian. \\

In this work, we define the correlation self-energy from $G_{\text{RS}}W_{\text{RS}}$ as the RSc correction, 
which means $\Delta^{\text{c}} = \Sigma^{G_{\text{RS}}W_{\text{RS}}}_{\text{c}}$.
In principle, 
this RSc scheme can be applied in other Green's function methods.
For example, $\Delta^{\text{c}}$ can be obtained from the correlation part of the T-matrix self-energy for T-matrix methods.
The $\Delta^{\text{c}}$ can be used to correct the RS Green's function $G_{\text{RS}}$ in three schemes. \\

First, the full RS Hamiltonian is corrected, 
which means $H^{\text{RSc}}_{pq} = H^{\text{RS}}_{pq} + \Delta^{\text{c}}_{pq}$.
The off-diagonal elements of $\Delta^{\text{c}}_{pq}$ are evaluated in the same manner as the Faleev's approximation for the correlation self-energy in qs$GW$\cite{kaplanQuasiParticleSelfConsistentGW2016,faleevAllElectronSelfConsistentGW2004}
\begin{equation}
    \Delta^{\text{c}}_{pq} = \frac{1}{2} [ \Sigma^{c}_{pq} (\epsilon_p) + \Sigma^{c}_{pq} (\epsilon_q) ] \text{,}
\end{equation}
Second, only diagonal elements of the RS Hamiltonian are corrected, 
which means $\Sigma^{c}_{pq} = \delta_{pq} \Sigma^{c}_{pq}$. 
In these two schemes, 
the RSc eigenvalues are obtained by diagonalizing the RSc Hamiltonian.
Third, the RS eigenvalues are shifted by the RSc correction to get the RSc eigenvalues
\begin{equation}\label{eq:rsc_c}
    \epsilon^{\text{RSc}}_p = \epsilon^{\text{RS}}_p + \Delta^{\text{c}}_{pp} \text{.}
\end{equation}
As shown in Section~5 in the Supporting Information, 
these three schemes provide similar results.
In the present work, 
we use the third scheme considering the computational cost.\\

The RSc Green's function $G_{\text{RSc}}$ is diagonal in the orbital space
\begin{equation}\label{eq:rsc_green}
    [G_{\text{RSc}}]_{pq} (\omega) = \delta_{pq}
    \frac{1}{\omega - \epsilon_{p}^{\text{RSc}} +
    i\eta \text{sgn} (\epsilon_{p}^{\text{RSc}} - \mu)} \text{.}
\end{equation}
$G_{\text{RSc}}$ has the same form as the KS Green's function, 
which means that it can be directly used in the $G_0W_0$ routine. 
In the second-shot calculation, we use KS orbitals for simplicity. 
As shown in Section~6 in the Supporting Information, 
using RSc orbitals gives similar results as using KS orbitals.\\

In the second shot, 
the RSc Green's function defined in Eq.\ref{eq:rsc_green} is used in the regular $G_0W_0$ routine, 
which is denoted as $G_{\text{RSc}}W_{\text{RSc}}$. 
The QP energies are obtained by solving the QP equation
\begin{equation}
    \epsilon_p^{\text{QP}} = \langle \psi_p^0 | H^{\text{H}} +
    \Sigma^{\text{x}} +
    \Sigma^{\text{c},G_{\text{RSc}}W_{\text{RSc}}} (\epsilon_p^{\text{QP}}) | \psi_p^0 \rangle \text{,}
\end{equation}
where $H^{\text{H}}$ is the Hartree Hamiltonian. 
The $G_{\text{RSc}}W_{\text{RSc}}$ self-energy has a similar form as $G_0W_0$\\
\begin{equation}
        \Sigma^{c,G_{\text{RSc}}W_{\text{RSc}}}_{pq} (\omega) = 
        \sum_{m} \sum_{r} \frac{ \langle \psi^0_p \psi^0_{r} | \rho^{\text{RSc}}_m \rangle \langle \psi^0_q \psi^0_{r} | \rho^{\text{RSc}}_m \rangle }{\omega - \epsilon^{\text{RSc}}_{r}
        - (\Omega^{\text{RSc}}_m - i\eta) \text{sgn} (\epsilon^{\text{RSc}}_{r} - \mu)} \text{, }
    \label{eq:grscwrsc_se}
\end{equation}
where $\rho^{\text{RSc}}_m$ and $\Omega^{\text{RSc}}_m$ are the transition densities
and the corresponding excitation energies from RPA calculated with the RSc Green's function. \\

The $G_{\text{RSc}}W_0$ method can be obtained by fixing the screened interaction at the DFT level,
which means the self-energy becomes
\begin{equation}
        \Sigma^{c,G_{\text{RSc}}W_0}_{pq} (\omega) = 
        \sum_{m} \sum_{r} \frac{ \langle \psi^0_p \psi^0_{r} | \rho^{0}_m \rangle \langle \psi^0_q \psi^0_{r} | \rho^{0}_m \rangle }{\omega - \epsilon^{\text{RSc}}_{r}
        - (\Omega^{0}_m - i\eta) \text{sgn} (\epsilon^{\text{RSc}}_{r} - \mu)} \text{, }
    \label{eq:grscw0_se}
\end{equation}

The QP orbitals from $G_{\text{RSc}}W_0$ and $G_{\text{RSc}}W_{\text{RSc}}$ can be obtained from diagonalizing the $GW$ Hamiltonian
\begin{equation}\label{eq:gw_hamiltonian}
    H^{GW} = H^{\text{H}} + \Sigma^{\text{x}} + \Sigma^{c} \text{,}
\end{equation}
where the last term $\Sigma^{c}$ in Eq.\ref{eq:gw_hamiltonian} is the correlation self-energy from the corresponding $GW$ method. \\

We implemented $G_{\text{RSc}}W_0$ and $G_{\text{RSc}}W_{\text{RSc}}$ methods in the QM4D quantum chemistry package\cite{qm4d}. 
We applied $GW$ methods including $G_{\text{RSc}}W_0$, $G_{\text{RSc}}W_{\text{RSc}}$, $G_{\text{RS}}W_0$, $G_{\text{RS}}W_{\text{RS}}$, $G_0W_0$, ev$GW_0$ and ev$GW$ to calculate IPs for the GW100 set\cite{vansettenGW100BenchmarkingG0W02015}, CLBEs and $\Delta$CLBEs for the CORE65 set\cite{golzeAccurateAbsoluteRelative2020}, and dipole moments of molecular systems\cite{kaplanQuasiParticleSelfConsistentGW2016}.
For the GW100 set,
the def2-TZVPP basis set was used\cite{weigendBalancedBasisSets2005}.
Reference values and geometries were taken from Ref.\citenum{vansettenGW100BenchmarkingG0W02015}.
Systems containing Xe, Rb, I, Ag and Cu were excluded because of the convergence problem in our program.
For the CORE65 set,
the cc-pVTZ basis set was used\cite{dunningGaussianBasisSets1989}.
Reference values and geometries were taken from Ref.\citenum{golzeAccurateAbsoluteRelative2020}. 
For dipole moment calculations, 
the def2-TZVPP basis set was used\cite{weigendBalancedBasisSets2005}.
Geometries and reference values of LiH, HF, LiF and CO were taken from Ref.\citenum{kaplanQuasiParticleSelfConsistentGW2016}. 
All the calculations were performed with QM4D\cite{qm4d}.
In QM4D, the $GW$ self-energy integral is calculated with the fully analytical treatment. 
For the calculations for core states, the QP equation is solved iteratively.
QM4D uses Cartesian basis sets and the resolution of identity (RI) technique\cite{weigendAccurateCoulombfittingBasis2006,renResolutionofidentityApproachHartree2012,eichkornAuxiliaryBasisSets1995}
to compute two-electron integrals.
All basis sets were taken from the Basis Set Exchange\cite{fellerRoleDatabasesSupport1996,pressNumericalRecipes2nd1992,schuchardtBasisSetExchange2007}. \\

\FloatBarrier

\begin{table}
\global\long\def\arraystretch{1.5}
\setlength{\tabcolsep}{7pt}
\caption{\label{tab:gw100}Mean absolute errors (MAEs) and mean signed errors (MSEs) of ionization potentials of the GW100 set obtained from $G_0W_0$, $G_{\text{RS}}W_0$, $G_{\text{RS}}W_{\text{RS}}$, ev$GW_0$, ev$GW$, $G_{\text{RSc}}W_0$ and $G_{\text{RSc}}W_{\text{RSc}}$ based on HF, BLYP, PBE, B3LYP and PBE0.
Geometries and reference values were taken from Ref.\citenum{vansettenGW100BenchmarkingG0W02015}.
The def2-TZVPP basis set was used.
All values in \,{eV}.
}
\begin{tabular}{c|cccccccccc}
\toprule
                                  & \multicolumn{2}{c}{HF} & \multicolumn{2}{c}{BLYP} & \multicolumn{2}{c}{PBE} & \multicolumn{2}{c}{B3LYP} & \multicolumn{2}{c}{PBE0} \\
                                  \cmidrule(l{0.5em}r{0.5em}){2-3} \cmidrule(l{0.5em}r{0.5em}){4-5} \cmidrule(l{0.5em}r{0.5em}){6-7} \cmidrule(l{0.5em}r{0.5em}){8-9} \cmidrule(l{0.5em}r{0.5em}){10-11}
                                  & MAE  & MSE  & MAE  & MSE   & MAE  & MSE   & MAE  & MSE    & MAE  & MSE \\
\midrule
$G_0W_0$                          & 0.50 & 0.20 & 0.69 & -0.63 & 0.67 & -0.61 & 0.49  & -0.37 & 0.45 & -0.33 \\
ev$GW_0$                          & 0.51 & 0.20 & 0.44 & -0.30 & 0.44 & -0.31 & 0.38  & -0.21 & 0.37 & -0.19 \\
ev$GW$                            & 0.49 & 0.16 & 0.35 & 0.04  & 0.36 & 0.03  & 0.33  & 0.01  & 0.31 & 0.03  \\
$G_{\text{RS}}W_0$                & 0.50 & 0.20 & 0.37 & -0.19 & 0.38 & -0.20 & 0.35  & -0.15 & 0.35 & -0.14 \\
$G_{\text{RS}}W_{\text{RS}}$      & 0.50 & 0.20 & 0.38 & 0.11  & 0.39 & 0.10  & 0.36  & 0.07  & 0.37 & 0.06  \\
$G_{\text{RSc}}W_0$               & 0.51 & 0.20 & 0.38 & -0.22 & 0.38 & -0.22 & 0.35  & -0.16 & 0.34 & -0.15 \\
$G_{\text{RSc}}W_{\text{RSc}}$    & 0.49 & 0.17 & 0.35 & 0.08  & 0.37 & 0.07  & 0.34  & 0.04  & 0.34 & 0.03  \\
\bottomrule
\end{tabular}
\end{table}

We first examine the performance of $G_{\text{RSc}}W_0$ and $G_{\text{RSc}}W_{\text{RSc}}$ on predicting valence QP energies.
The mean absolute errors (MAEs) and mean signed errors (MSEs) of calculated IPs of the GW100 set obtained from various $GW$ methods including $G_0W_0$, $G_{\text{RS}}W_0$, $G_{\text{RS}}W_{\text{RS}}$, ev$GW_0$,
ev$GW$, $G_{\text{RSc}}W_0$ and $G_{\text{RSc}}W_{\text{RSc}}$ based on HF, BLYP, PBE, B3LYP and PBE0 are tabulated in Table.\ref{tab:gw100}. 
It is well-known that the perturbative $G_0W_0$ method has the dependence on the choice of the DFA.
This dependence can be attributed to the underscreening error and the overscreening error\cite{golzeGWCompendiumPractical2019,carusoSelfconsistentGWAllelectron2013,maromBenchmarkGWMethods2012},
consequence of the localization error and the delocalization error of DFAs\cite{mori-sanchezDiscontinuousNatureExchangeCorrelation2009,cohenInsightsCurrentLimitations2008}. 
The HF starting point typically overestimates the fundamental gap that is inversely proportional to the response function.
This leads to the underestimated screened interaction, 
which is known as the underscreening error\cite{golzeGWCompendiumPractical2019,carusoBenchmarkGWApproaches2016,maromBenchmarkGWMethods2012}.
Similarly, 
commonly used DFAs underestimate the fundamental gap and lead to the overscreening error. 
Thus, $G_0W_0$ based on GGA or hybrid functionals underestimates IPs with MSEs around $0.3$ \,{eV} to $0.7$ \,{eV} and $G_0W_0$ based on HF overestimates IPs with a MSE around $0.2$ \,{eV}. 
The underscreening error can be largely reduced by introducing the self-consistency in $GW$ calculations, 
because fundamental gaps from the QP energies at the $GW$ level are generally larger than those from KS eigenvalues.
As shown in Table.\ref{tab:gw100},
ev$GW$ significantly outperforms $G_0W_0$ and greatly washes out the starting point dependence. 
ev$GW$ with commonly used DFAs only slightly overestimates IPs with MSEs around $0.03$ \,{eV}. 
ev$GW$@HF has relatively larger errors than ev$GW$ based on other DFAs because HF orbitals lack the description for the dynamical correlation of the system.
As expected, ev$GW_0$ has a larger starting point dependence than ev$GW$, 
owing to the non-self-consistent treatment for the screened interaction. 
Our results show that ev$GW_0$ with hybrid functionals predicts more accurate IPs of the GW100 set than ev$GW_0$ with GGA functionals,
which is similar to the results for sc$GW_0$ reported in Ref.\citenum{carusoBenchmarkGWApproaches2016}.
However, 
this "best $W$" strategy\cite{martinInteractingElectrons2016,golzeGWCompendiumPractical2019} depends on the systems and the properties of interest.
As shown in Ref.\citenum{maromBenchmarkGWMethods2012} and Ref.\citenum{golzeAccurateAbsoluteRelative2020}, 
fixing the screened interaction at the PBE level is optimal for sc$GW_0$ to predict IPs of azabenzenes and for ev$GW_0$ to predict CLBEs of the CORE65 set. 
$G_{\text{RS}}W_0$ and $G_{\text{RS}}W_{\text{RS}}$ based on HF are equivalent to $G_0W_0$@HF as expected. 
The starting point dependence in $G_{\text{RS}}W_0$ and $G_{\text{RS}}W_{\text{RS}}$ is largely reduced because the singles contributions are completely included.
The MAEs of $G_{\text{RS}}W_0$ and $G_{\text{RS}}W_{\text{RS}}$ are similar. 
As shown in Table.\ref{tab:gw100} $G_{\text{RS}}W_{\text{RS}}$ gives slightly larger MAEs and MSEs than ev$GW$, 
because the screening interaction calculated with the RS Green's function is underscreened.
The fundamental gap is properly reduced by using the RSc Green's function. 
$G_{\text{RSc}}W_{\text{RSc}}$ provides very similar results as ev$GW$ and the starting point dependence is largely eliminated. 
The difference of using different DFAs in $G_{\text{RSc}}W_{\text{RSc}}$ is smaller than $0.05$ \,{eV}.
We also found that $G_{\text{RSc}}W_0$ and $G_{\text{RSc}}W_{\text{RSc}}$ has similar accuracy to ev$GW_0$ and ev$GW$, respectively. 

\FloatBarrier

\begin{figure}
\includegraphics[width=0.6\textwidth]{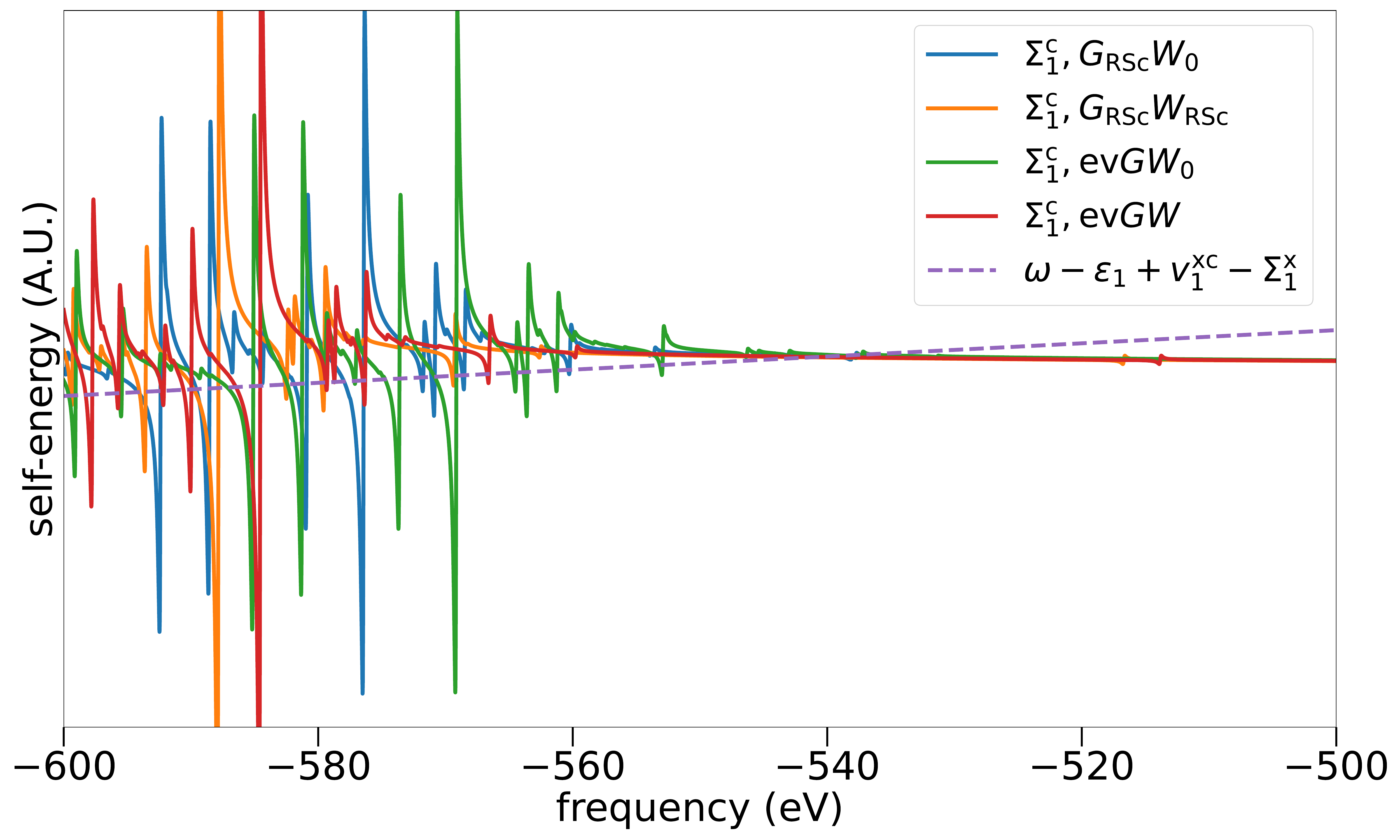}
\caption{Graphical solutions of $\text{O}_{\text{1s}}$ excitation in the water
molecule from ev$GW_0$, ev$GW$, $G_{\text{RSc}}W_0$ and $G_{\text{RSc}}W_{\text{RSc}}$ based on PBE.
The graphical solutions are found at intersections between $\omega-\epsilon_{\text{1s}}+v_{\text{1s}}^{\text{xc}}-\Sigma_{\text{1s}}^{\text{x}}$
(the dashed line) and the correlation part of $\text{O}_{\text{1s}}$
self-energy. 
The geometry was taken from Ref.\citenum{golzeAccurateAbsoluteRelative2020}.
The cc-pVTZ basis set was used.}
\label{fig:graphical_solution} 
\end{figure}

\begin{figure}
\includegraphics[width=0.6\textwidth]{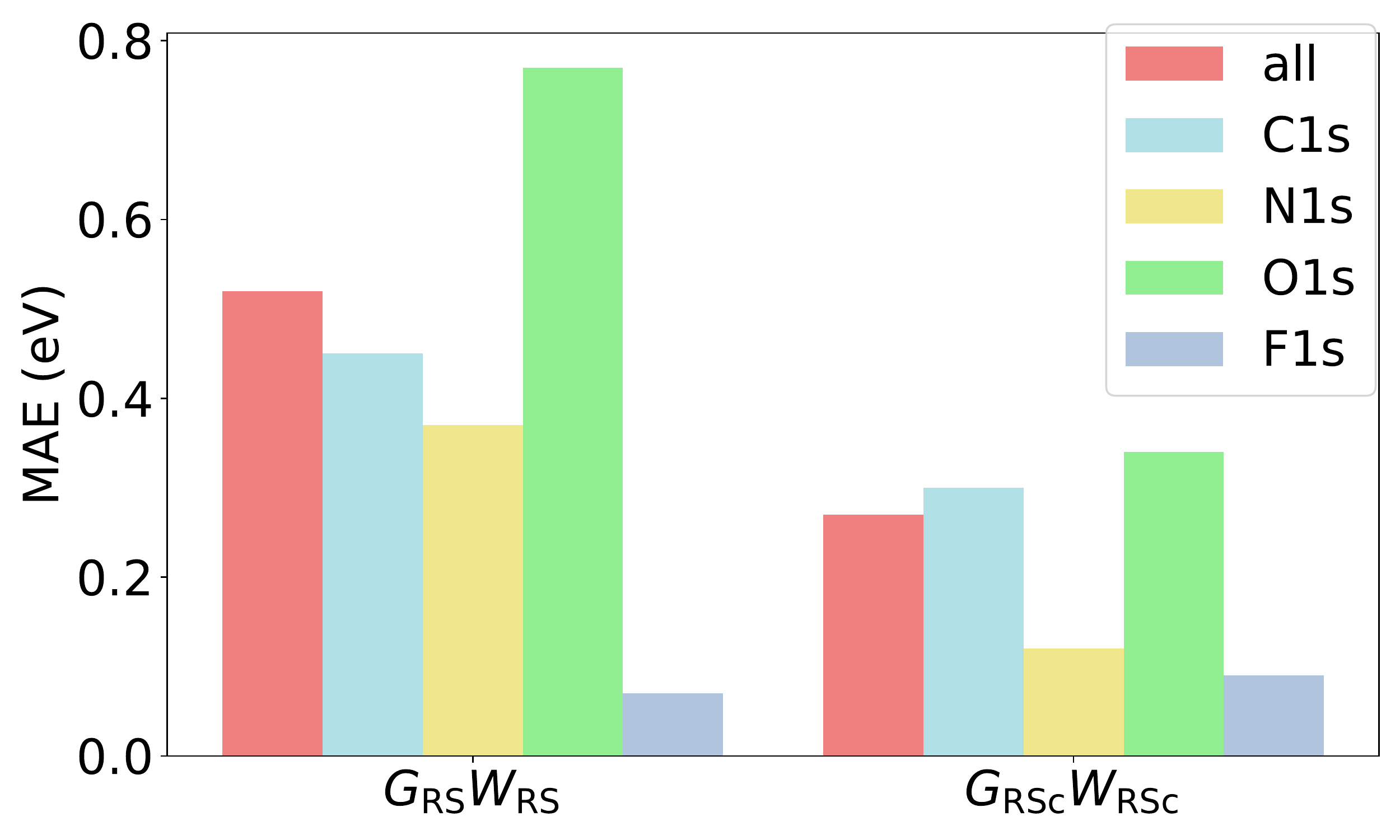}
\caption{Mean absolute errors of calculated relative core-level binding energies for different species obtained from $G_{\text{RS}}W_{\text{RS}}$@PBE and $G_{\text{RSc}}W_{\text{RSc}}$@PBE.}
\label{fig:relative_bar} 
\end{figure}

\begin{table}
\global\long\def\arraystretch{1.5}
\setlength{\tabcolsep}{7pt}
\caption{\label{tab:core65_absolute}Mean absolute errors (MAEs) and mean signed errors (MSEs) of core-level binding energies of the CORE65 set obtained from $G_0W_0$, $G_{\text{RS}}W_0$, $G_{\text{RS}}W_{\text{RS}}$, ev$GW_0$, ev$GW$, $G_{\text{RSc}}W_0$ and $G_{\text{RSc}}W_{\text{RSc}}$ based on HF, BLYP, PBE, B3LYP and PBE0.
Geometries and reference values were taken from Ref.\citenum{golzeAccurateAbsoluteRelative2020}.
The cc-pVTZ basis set was used.
All values in \,{eV}.
}
\begin{tabular}{c|cccccccccc}
\toprule
                                  & \multicolumn{2}{c}{HF} & \multicolumn{2}{c}{BLYP} & \multicolumn{2}{c}{PBE} & \multicolumn{2}{c}{B3LYP} & \multicolumn{2}{c}{PBE0} \\
                                  \cmidrule(l{0.5em}r{0.5em}){2-3} \cmidrule(l{0.5em}r{0.5em}){4-5} \cmidrule(l{0.5em}r{0.5em}){6-7} \cmidrule(l{0.5em}r{0.5em}){8-9} \cmidrule(l{0.5em}r{0.5em}){10-11}
                                  & MAE  & MSE  & MAE  & MSE   & MAE  & MSE   & MAE  & MSE    & MAE  & MSE \\
\midrule
$G_0W_0$                          & 5.66 & 5.66 &      &       &      &       &       &       &      &       \\
ev$GW_0$                          & 3.61 & 3.61 & 1.81 & -1.81 & 1.50 & -1.50 & 0.65  & -0.65 & 0.26 & -0.08 \\
ev$GW$                            & 3.00 & 3.00 & 0.56 & 0.56  & 0.89 & 0.89  & 0.98  & 0.98  & 1.39 & 1.39  \\
$G_{\text{RS}}W_0$                & 5.66 & 5.66 &      &       &      &       &       &       &      &       \\
$G_{\text{RS}}W_{\text{RS}}$      & 5.66 & 5.66 & 3.52 & 3.52  & 3.81 & 3.81  & 3.87  & 3.87  & 4.21 & 4.21  \\
$G_{\text{RSc}}W_0$               & 4.02 & 4.02 & 0.40 & -0.04 & 0.58 & 0.33  & 0.62  & 0.62  & 1.13 & 1.13  \\
$G_{\text{RSc}}W_{\text{RSc}}$    & 3.63 & 3.63 & 1.21 & 1.21  & 1.53 & 1.53  & 1.63  & 1.63  & 2.02 & 2.02  \\
\bottomrule
\end{tabular}
\end{table}

\begin{table}
\global\long\def\arraystretch{1.5}
\setlength{\tabcolsep}{7pt}
\caption{\label{tab:core65_relative}Mean absolute errors (MAEs) and mean signed errors (MSEs) of relative core-level binding energies ($\Delta$CLBEs) of the CORE65 set obtained from $G_0W_0$, $G_{\text{RS}}W_0$, $G_{\text{RS}}W_{\text{RS}}$, ev$GW_0$, ev$GW$, $G_{\text{RSc}}W_0$ and $G_{\text{RSc}}W_{\text{RSc}}$ based on HF, BLYP, PBE, B3LYP and PBE0. 
The $\Delta$CLBEs are the shifts with respect to a reference molecule, $\Delta\textnormal{CLBE}=\textnormal{CLBE}-\textnormal{CLBE}_{\textnormal{ref\_mol}}$. \ce{CH4}, \ce{NH3}, \ce{H2O} and \ce{CH3F} have been used as reference molecules for C1s, N1s, O1s and F1s respectively.
Geometries and reference values were taken from Ref.\citenum{golzeAccurateAbsoluteRelative2020}.
The cc-pVTZ basis set was used.
All values in \,{eV}.
}
\begin{tabular}{c|cccccccccc}
\toprule
                                  & \multicolumn{2}{c}{HF} & \multicolumn{2}{c}{BLYP} & \multicolumn{2}{c}{PBE} & \multicolumn{2}{c}{B3LYP} & \multicolumn{2}{c}{PBE0} \\
                                  \cmidrule(l{0.5em}r{0.5em}){2-3} \cmidrule(l{0.5em}r{0.5em}){4-5} \cmidrule(l{0.5em}r{0.5em}){6-7} \cmidrule(l{0.5em}r{0.5em}){8-9} \cmidrule(l{0.5em}r{0.5em}){10-11}
                                  & MAE  & MSE  & MAE  & MSE   & MAE  & MSE   & MAE  & MSE    & MAE  & MSE \\
\midrule
$G_0W_0$                          & 0.67 & 0.65 &      &       &      &       &      &      &      &      \\
ev$GW_0$                          & 0.51 & 0.40 & 0.49 & 0.04  & 0.66 & 0.25  & 0.21 & 0.08 & 0.22 & 0.01 \\
ev$GW$                            & 0.30 & 0.27 & 0.19 & -0.09 & 0.20 & -0.07 & 0.16 & 0.06 & 0.17 & 0.08 \\
$G_{\text{RS}}W_0$                & 0.67 & 0.65 &      &       &      &       &      &      &      &      \\
$G_{\text{RS}}W_{\text{RS}}$      & 0.67 & 0.65 & 0.47 & 0.29  & 0.52 & 0.30  & 0.48 & 0.39 & 0.51 & 0.41 \\
$G_{\text{RSc}}W_0$               & 0.51 & 0.50 & 0.36 & 0.10  & 0.44 & 0.17  & 0.38 & 0.34 & 0.41 & 0.35 \\
$G_{\text{RSc}}W_{\text{RSc}}$    & 0.39 & 0.38 & 0.23 & 0.06  & 0.27 & 0.08  & 0.25 & 0.19 & 0.28 & 0.21 \\
\bottomrule
\end{tabular}
\end{table}

\FloatBarrier

We then examine the performance of $G_{\text{RSc}}W_0$ and $G_{\text{RSc}}W_{\text{RSc}}$ on predicting core-level QP energies.
As discussed in literature\cite{golzeCoreLevelBindingEnergies2018,golzeAccurateAbsoluteRelative2020,moninoUnphysicalDiscontinuitiesIntruder2022,disabatinoScrutinizingGWBasedMethods2021}, 
the multi-solution issue is frequently encountered in $GW$ methods, especially for core-level state calculations.
The solution behaviors of $G_0W_0$, ev$GW_0$, ev$GW$, $G_{\text{RS}}W_0$ and $G_{\text{RS}}W_{\text{RS}}$ were studied in previous works\cite{golzeAccurateAbsoluteRelative2020,golzeCoreLevelBindingEnergies2018,li2022benchmark}. \\

The solution behaviors of $G_{\text{RSc}}W_0$ and $G_{\text{RSc}}W_{\text{RSc}}$ are shown in Fig.\ref{fig:graphical_solution}.
By checking the graphical solution,
both $G_{\text{RSc}}W_0$ and $G_{\text{RSc}}W_{\text{RSc}}$ show a unique solution corresponding to the desired QP state. 
In $G_{\text{RSc}}W_0$, using the RSc eigenvalues in the Green's function shifts poles of satellites to the negative direction (away from the main peak). 
For $G_{\text{RSc}}W_{\text{RSc}}$, RSc eigenvalues are also in the screened interaction, 
which leads to larger RPA excitation energies and a stronger shift. 
As shown in Fig.\ref{fig:graphical_solution},
the self-energy of $G_{\text{RSc}}W_0$ and $G_{\text{RSc}}W_{\text{RSc}}$ is analogous to ev$GW_0$ and and ev$GW$, respectively. 
$G_{\text{RSc}}W_{\text{RSc}}$ and ev$GW$ provide stronger shifts than $G_{\text{RSc}}W_0$ and ev$GW_0$. \\

The MAEs and MSEs of calculated CLBEs and $\Delta$CLBEs of the CORE65 set obtained from various $GW$ methods including $G_0W_0$, $G_{\text{RS}}W_0$, $G_{\text{RS}}W_{\text{RS}}$, ev$GW_0$,
ev$GW$, $G_{\text{RSc}}W_0$ and $G_{\text{RSc}}W_{\text{RSc}}$ based on HF, BLYP, PBE, B3LYP and PBE0 are tabulated in Table.\ref{tab:core65_absolute} and Table.\ref{tab:core65_relative}, respectively.
We start with the discussion for CLBEs. 
As shown in Table.\ref{tab:core65_absolute} and our previous work\cite{li2022benchmark}, 
although $G_{\text{RS}}W_{\text{RS}}$ can provide a unique solution to the QP equation for core states, 
it overestimates CLBEs due to the underscreening error\cite{li2022benchmark}.
The underscreening error is greatly reduced by adding the correlation correction to RS eigenvalues,
which properly reduces the fundamental gap.
$G_{\text{RSc}}W_{\text{RSc}}$ significantly outperforms $G_{\text{RS}}W_{\text{RS}}$ for predicting CLBEs with MAEs around $1.5$ \,{eV}.
Both ev$GW$ and $G_{\text{RSc}}W_{\text{RSc}}$ overestimate CLBEs, 
which agrees with the stronger shift in the self-energy as shown in Fig.\ref{fig:graphical_solution}.
The systematical overestimation in $G_{\text{RSc}}W_{\text{RSc}}$ stems from the underscreening error in the $GW$ formalism, 
which needs to be compensated by the vertex correction\cite{golzeAccurateAbsoluteRelative2020,golzeGWCompendiumPractical2019}. 
Similar to the "best $W$" strategy used in ev$GW_0$\cite{martinInteractingElectrons2016,golzeAccurateAbsoluteRelative2020,golzeGWCompendiumPractical2019}, 
the underscreening error in $G_{\text{RSc}}W_{\text{RSc}}$ can be largely compensated by fixing the screened interaction at the DFT level.
Because KS eigenvalues provide the underestimated fundamental gap that leads to an overscreened interaction,
the error compensation in $G_{\text{RSc}}W_0$ leads to the improved accuracy. 
We found that $G_{\text{RSc}}W_0$ based on GGA functionals provides better accuracy with MAEs only around $0.5$ \,{eV}. 
As shown in Table.\ref{tab:core65_absolute} ev$GW_0$ based on hybrid functionals provides better accuracy than ev$GW_0$ based on GGA functionals, 
which is different from the conclusion in Ref.\citenum{golzeAccurateAbsoluteRelative2020} and our previous work\cite{li2022benchmark}.
The CLBEs in the present work were calculated with the cc-pVTZ basis set,
which are slightly underestimated. 
Extrapolating the results to the complete basis set limit and adding the relativistic correction slightly increase calculated CLBEs by $0.3$ \,{eV}\cite{li2022benchmark}, 
which improve the accuracy of ev$GW_0$ with GGA functionals. 
Because CLBEs obtained from $G_{\text{RSc}}W_0$ with GGA functionals are smaller than those obtained from $G_{\text{RSc}}W_0$ with hybrid functionals, 
we do not expect the extrapolation and the relativistic correction to change our conclusion that $G_{\text{RSc}}W_0$ with GGA functionals can predict accurate CLBEs.\\ 

We move to the discussion for $\Delta$CLBEs.
In practice, the chemical shifts between CLBEs for the same specie in different chemical environments can be smaller than $0.5$ \,{eV}\cite{siegbahnESCAAppliedFree1970}.
Thus, $G_{\text{RSc}}W_{\text{RSc}}$ that gives MAEs around $0.3$ \,{eV} is promising for predicting $\Delta$CLBEs.
The accuracy of $G_{\text{RSc}}W_0$ is comparable to ev$GW_0$.
$G_{\text{RSc}}W_0$ provides slightly larger MAEs around $0.4$ \,{eV}.
As shown in Ref.\citenum{li2022benchmark}, 
$G_{\text{RS}}W_{\text{RS}}$ has a strong specie dependence for predicting $\Delta$CLBEs, 
which originates from the DFT calculations.
This dependence is largely reduced in $G_{\text{RSc}}W_{\text{RSc}}$. 
The MAEs of calculated $\Delta$CLBEs obtained from $G_{\text{RS}}W_{\text{RS}}$@PBE and $G_{\text{RSc}}W_{\text{RSc}}$@PBE for different species are plotted in Fig.\ref{fig:relative_bar}.
The MAEs of for $G_{\text{RSc}}W_{\text{RSc}}$@PBE predicting $\Delta$CLBEs of all species are largely reduced. 
The specie dependence is only around $0.15$ \,{eV}.\\

\FloatBarrier

\begin{table}[]
\global\long\def\arraystretch{1.5}
\setlength{\tabcolsep}{15pt}
    \caption{\label{tab:dipole}Mean absolute errors of calculated dipole moments of LiH, HF, LiF and CO obtained from DFT, $G_0W_0$, $G_{\text{RS}}W_0$, $G_{\text{RS}}W_{\text{RS}}$, $G_{\text{RSc}}W_0$ and $G_{\text{RSc}}W_{\text{RSc}}$ with HF, BLYP, PBE, B3LYP and PBE0.
    Geometries, reference values and qs$GW$ results were taken from Ref.\citenum{kaplanQuasiParticleSelfConsistentGW2016}.
    The def2-TZVPP basis set was used.
    All values in Debye.}
    \begin{tabular}{c|ccccc}
            \hline
            & HF   & BLYP & PBE  & B3LYP & PBE0 \\
            \hline
    KS                             & 0.36 & 0.22 & 0.19 & 0.18  & 0.16 \\
    $G_0W_0$                       & 0.15 & 0.28 & 0.25 & 0.16  & 0.06 \\
    $G_{\text{RS}}W_0$             & 0.15 & 0.13 & 0.12 & 0.23  & 0.15 \\
    $G_{\text{RS}}W_{\text{RS}}$   & 0.15 & 0.11 & 0.11 & 0.06  & 0.13 \\
    $G_{\text{RSc}}W_0$            & 0.09 & 0.17 & 0.18 & 0.12  & 0.11 \\
    $G_{\text{RSc}}W_{\text{RSc}}$ & 0.06 & 0.08 & 0.10 & 0.09  & 0.08 \\
    qs$GW$                         & 0.03 &      &      &       &      \\
    \hline
    \end{tabular}
\end{table}

In addition to the improved accuracy for predicting QP energies,
the orbital update from $G_{\text{RSc}}W_0$ and $G_{\text{RSc}}W_{\text{RSc}}$ can be obtained by constructing and diagonalizing the $GW$ Hamiltonian defined in Eq.\ref{eq:gw_hamiltonian}.
The orbital update from other $GW$ methods, for example $G_0W_0$, can be obtained in the same manner.
We investigate the performance of $G_{\text{RSc}}W_0$ and $G_{\text{RSc}}W_{\text{RSc}}$ on updating QP orbitals by calculating dipole moments of four molecules LiH, HF, LiF and CO.
The MAEs of calculated dipole moments obtained from DFT, $G_0W_0$, $G_{\text{RS}}W_0$, $G_{\text{RS}}W_{\text{RS}}$, $G_{\text{RSc}}W_0$ and $G_{\text{RSc}}W_{\text{RSc}}$ based on HF, BLYP, PBE, B3LYP and PBE0 are tabulated in Table.\ref{tab:dipole}. 
It can be seen that dipole moments predicted by DFT calculations depend on the choice of the DFA and the errors are relatively large.
It has been shown in Ref.\citenum{liCombiningLocalizedOrbital2022} that using localized orbitals provides improved accuracy for predicting dipole moments of small molecules.
As shown in Ref.\citenum{kaplanQuasiParticleSelfConsistentGW2016} and Ref.\citenum{carusoBenchmarkGWApproaches2016},
qs$GW$ that also provides more localized orbitals outperforms DFAs for predicting dipole moments.
The $G_{\text{RSc}}W_{\text{RSc}}$ method provides a similar accuracy to qs$GW$ with all tested DFAs,
which means the starting point dependence in QP orbitals is largely eliminated in an efficient manner and the quality of $G_{\text{RSc}}W_{\text{RSc}}$ orbitals is close to qs$GW$. \\

We developed the RSc Green's function in the $GW$ formalism for predicting accurate QP energies and orbitals. 
The RSc Green's function captures all exchange contributions from the associated DFA and treats correlation contributions in a perturbative manner.
Two methods were introduced:
$G_{\text{RSc}}W_{\text{RSc}}$ that uses the RSc Green's function as the starting point and formulates the screened interaction with the RSc Green's function,
$G_{\text{RSc}}W_0$ that uses the RSc Green's function as the starting point and fixes the screened interaction at the DFA level.
These two methods were first examined on valence states by calculating IPs of the GW100 set. 
$G_{\text{RSc}}W_{\text{RSc}}$ largely reduces the starting point dependence and provides the similar accuracy to ev$GW$.
Then we studied CLBEs in the CORE65 set.
$G_{\text{RSc}}W_{\text{RSc}}$ slightly overestimates the results because of the underscreening error, 
which is similar to ev$GW$. 
By applying the "best $W$" strategy used in ev$GW_0$, 
$G_{\text{RSc}}W_0$ with GGA functionals predicts accurate CLBEs. 
We also studied the QP orbitals from $G_{\text{RSc}}W_{\text{RSc}}$ and $G_{\text{RSc}}W_0$. 
The orbital update from these two methods is obtained from the one-time diagonalization of the $GW$ Hamiltonian.
We showed that $G_{\text{RSc}}W_{\text{RSc}}$ predicts accurate dipole moments of small molecules.
$G_{\text{RSc}}W_{\text{RSc}}$ and $G_{\text{RSc}}W_0$ are twice the computation cost of $G_0W_0$,
which is much more favorable than all other self-consistent $GW$ methods. 
Thus, we demonstrate an efficient alternative to iterating Hedin's equations for obtain accurate QP energies and orbitals.
The RSc Green's function can be applied to other Green's function methods,
which is promising for making the Green's function approach more efficient and robust.\\

\begin{acknowledgement}
ACKNOWLEDGMENTS: J. L. acknowledges the support from the National
Institute of General Medical Sciences of the National Institutes of
Health under award number R01-GM061870. W.Y. acknowledges the support
from the National Science Foundation (grant no. CHE-1900338).
\end{acknowledgement}
\begin{suppinfo}
Supporting Information Available: formulation of the renormalized singles Green's function, ionization potentials of the GW100 set, core-level binding energies of the CORE65 set, dipole moments of small molecules, comparison of different schemes of the RSc correction, comparison of using RSc orbitals and KS orbitals in $G_{\text{RSc}}W_{\text{RSc}}$.
\end{suppinfo}

\bibliography{ref,software}

\end{document}


\section{RS Green's Function}
To use the form of the HF self-energy to eliminate the starting point dependence, 
the RS Green's function is defined as
\begin{equation}
    G_{\text{RS}}^{-1} = G_0^{-1} - P(\Sigma_{\text{Hx}}[G_{0}]-v_{\text{Hxc}})P - Q(\Sigma_{\text{Hx}}[G_{0}]-v_{\text{Hxc}})Q \text{,}
\end{equation}
where $P=\sum_{i}^{occ}|\psi^{(0)}_{i}\rangle\langle\psi^{(0)}_{i}|$ is the projection into the occupied orbital space and $Q=I-P$ is the projection into the virtual orbital space,
$\Sigma_{\text{Hx}}[G_{0}]$ means that the HF (Hartree and exchange) self-energy is constructed with the KS Green's function, 
$\{ \psi^0 \}$ are KS orbitals.
We use $i$, $j$ for occupied orbitals, $a$, $b$ for virtual orbitals, $p$, $q$ for general orbitals.\\

Equivalently, the RS Green's function is obtained by using the KS density matrix in the HF Hamiltonian,
which is the RS Hamiltonian $H^{\text{RS}} = H^{\text{HF}}[G_{0}]$,
and solving two projected HF equations in the occupied subspace and the virtual subspaces\cite{jinRenormalizedSinglesGreen2019}
\begin{equation}
    P(H^{\text{HF}}[G_{0}])P|\psi_{i}^{\text{RS}}\rangle =
    P ( H^{\text{RS}} ) P | \psi_{i}^{\text{RS}} \rangle =
    \epsilon_{i}^{\text{RS}}P|\psi_{i}^{\text{RS}}\rangle \text{,}\label{eq:rs_occ}
\end{equation}
and
\begin{equation}
    Q(H^{\text{HF}}[G_{0}])Q|\psi_{a}^{\text{RS}}\rangle =
    Q ( H^{\text{RS}} ) Q | \psi_{i}^{\text{RS}} \rangle =
    \epsilon_{a}^{\text{RS}}Q|\psi_{a}^{\text{RS}}\rangle \text{.}\label{eq:rs_vir}
\end{equation}

\section{Ionization Potentials of the GW100 Set Obtained from Different $GW$ Methods}

\fontsize{9}{11}\selectfont{
\begin{longtable}{@{\extracolsep{\fill}}cccccccc}
\caption{Ionization potentials of the GW100 set\cite{vansettenGW100BenchmarkingG0W02015} obtained from $G_{0}W_{0}$ with HF, BLYP, PBE, B3LYP and PBE0. 
The def2-TZVPP basis set was used.
Systems containing Xe, Rb, I, Ag and Cu were excluded because of the convergence problem.
All values in eV. }\\
\toprule
& & & \multicolumn{5}{c}{$G_{0}W_{0}$} \\
\cmidrule(l{0.5em}r{0.5em}){4-8} 
name & formula & Exp & HF & BLYP & PBE & B3LYP & PBE0 \\
\hline 
\endfirsthead
\caption{Continued}\\
\toprule
& & & \multicolumn{5}{c}{$G_{0}W_{0}$} \\
\cmidrule(l{0.5em}r{0.5em}){4-8} 
name & formula & Exp & HF & BLYP & PBE & B3LYP & PBE0 \\
\hline
\endhead
\bottomrule
\endlastfoot
helium                 & \ce{He}       & 24.59 & 24.60 & 23.80 & 23.84 & 24.06 & 24.12 \\
neon                   & \ce{Ne}       & 21.56 & 21.41 & 20.54 & 20.57 & 20.79 & 20.85 \\
argon                  & \ce{Ar}       & 15.76 & 15.73 & 15.05 & 15.05 & 15.22 & 15.25 \\
krypton                & \ce{Kr}       & 14.00 & 13.96 & 13.36 & 13.35 & 13.50 & 13.52 \\
hydrogen               & \ce{H2}       & 15.43 & 16.49 & 15.93 & 15.96 & 16.14 & 16.19 \\
lithium dimer          & \ce{Li2}      & 4.73  & 5.29  & 5.02  & 5.08  & 5.17  & 5.22  \\
sodium dimer           & \ce{Na2}      & 4.89  & 4.95  & 4.85  & 4.88  & 4.94  & 4.95  \\
sodium tetramer        & \ce{Na4}      & 4.27  & 4.26  & 4.05  & 4.11  & 4.17  & 4.20  \\
\end{longtable}
}

\fontsize{9}{11}\selectfont{
\setlength{\tabcolsep}{4pt}
\begin{longtable}{@{\extracolsep{\fill}}ccccccccccccc}
\caption{Ionization potentials of the GW100 set\cite{vansettenGW100BenchmarkingG0W02015} obtained from ev$GW_0$ and ev$GW$ with HF, BLYP, PBE, B3LYP and PBE0. 
The def2-TZVPP basis set was used.
Systems containing Xe, Rb, I, Ag and Cu were excluded because of the convergence problem.
All values in eV. }\\
\toprule
& & & \multicolumn{5}{c}{ev$GW_0$} & \multicolumn{5}{c}{ev$GW$} \\
\cmidrule(l{0.5em}r{0.5em}){4-8} \cmidrule(l{0.5em}r{0.5em}){9-13}
name & formula & Exp & HF & BLYP & PBE & B3LYP & PBE0 & HF & BLYP & PBE & B3LYP & PBE0 \\
\hline 
\endfirsthead
\caption{Continued}\\
\toprule
& & & \multicolumn{5}{c}{ev$GW_0$} & \multicolumn{5}{c}{ev$GW$} \\
\cmidrule(l{0.5em}r{0.5em}){4-8} \cmidrule(l{0.5em}r{0.5em}){9-13}
name & formula & Exp & HF & BLYP & PBE & B3LYP & PBE0 & HF & BLYP & PBE & B3LYP & PBE0 \\
\hline
\endhead
\bottomrule
\endlastfoot
helium                 & \ce{He}       & 24.59 & 24.60 & 24.20 & 24.25 & 24.32 & 24.36 & 24.58 & 24.63 & 24.67 & 24.62 & 24.64  \\
neon                   & \ce{Ne}       & 21.56 & 21.33 & 21.09 & 21.12 & 21.12 & 21.15 & 21.25 & 21.71 & 21.72 & 21.56 & 21.54  \\
argon                  & \ce{Ar}       & 15.76 & 15.71 & 15.27 & 15.25 & 15.33 & 15.33 & 15.68 & 15.58 & 15.55 & 15.55 & 15.52  \\
krypton                & \ce{Kr}       & 14.00 & 13.94 & 13.55 & 13.52 & 13.59 & 13.58 & 13.92 & 13.81 & 13.77 & 13.77 & 13.74  \\
hydrogen               & \ce{H2}       & 15.43 & 16.51 & 16.26 & 16.28 & 16.34 & 16.37 & 16.51 & 16.52 & 16.53 & 16.52 & 16.52  \\
lithium dimer          & \ce{Li2}      & 4.73  & 5.34  & 5.20  & 5.22  & 5.23  & 5.26  & 5.34  & 5.32  & 5.33  & 5.29  & 5.31   \\
sodium dimer           & \ce{Na2}      & 4.89  & 4.99  & 4.97  & 4.98  & 4.98  & 4.99  & 5.00  & 5.03  & 5.04  & 5.01  & 5.01   \\
sodium tetramer        & \ce{Na4}      & 4.27  & 4.30  & 4.16  & 4.18  & 4.20  & 4.22  & 4.31  & 4.21  & 4.23  & 4.22  & 4.23   \\
\end{longtable}
}

\fontsize{9}{11}\selectfont{
\setlength{\tabcolsep}{4pt}
\begin{longtable}{@{\extracolsep{\fill}}ccccccccccccc}
\caption{Ionization potentials of the GW100 set\cite{vansettenGW100BenchmarkingG0W02015} obtained from $G_{\text{RS}}W_0$ and $G_{\text{RS}}W_{\text{RS}}$ with HF, BLYP, PBE, B3LYP and PBE0. 
The def2-TZVPP basis set was used.
Systems containing Xe, Rb, I, Ag and Cu were excluded because of the convergence problem.
All values in eV. }\\
\toprule
& & & \multicolumn{5}{c}{$G_{\text{RS}}W_0$} & \multicolumn{5}{c}{$G_{\text{RS}}W_{\text{RS}}$} \\
\cmidrule(l{0.5em}r{0.5em}){4-8} \cmidrule(l{0.5em}r{0.5em}){9-13} 
name & formula & Exp & HF & BLYP & PBE & B3LYP & PBE0 & HF & BLYP & PBE & B3LYP & PBE0 \\
\hline 
\endfirsthead
\caption{Continued}\\
\toprule
& & & \multicolumn{5}{c}{$G_{\text{RS}}W_0$} & \multicolumn{5}{c}{$G_{\text{RS}}W_{\text{RS}}$} \\
\cmidrule(l{0.5em}r{0.5em}){4-8} \cmidrule(l{0.5em}r{0.5em}){9-13} 
name & formula & Exp & HF & BLYP & PBE & B3LYP & PBE0 & HF & BLYP & PBE & B3LYP & PBE0 \\
\hline
\endhead
\bottomrule
\endlastfoot
helium                 & \ce{He}       & 24.59 & 24.60 & 24.29 & 24.34 & 24.37 & 24.41 &24.60 & 24.67 & 24.71 & 24.66 & 24.67 \\
neon                   & \ce{Ne}       & 21.56 & 21.41 & 21.34 & 21.36 & 21.31 & 21.32 &21.41 & 21.90 & 21.92 & 21.74 & 21.71 \\
argon                  & \ce{Ar}       & 15.76 & 15.73 & 15.37 & 15.35 & 15.40 & 15.39 &15.73 & 15.65 & 15.62 & 15.61 & 15.58 \\
krypton                & \ce{Kr}       & 14.00 & 13.96 & 13.65 & 13.62 & 13.66 & 13.64 &13.96 & 13.88 & 13.84 & 13.83 & 13.79 \\
hydrogen               & \ce{H2}       & 15.43 & 16.49 & 16.29 & 16.31 & 16.34 & 16.36 &16.49 & 16.51 & 16.53 & 16.50 & 16.51 \\
lithium dimer          & \ce{Li2}      & 4.73  & 5.29  & 5.15  & 5.17  & 5.17  & 5.20  &5.29  & 5.29  & 5.30  & 5.27  & 5.28  \\
sodium dimer           & \ce{Na2}      & 4.89  & 4.95  & 4.90  & 4.91  & 4.91  & 4.92  &4.95  & 4.98  & 4.98  & 4.95  & 4.95  \\
sodium tetramer        & \ce{Na4}      & 4.27  & 4.26  & 4.08  & 4.11  & 4.12  & 4.15  &4.26  & 4.15  & 4.17  & 4.16  & 4.17  \\
\end{longtable}
}

\fontsize{9}{11}\selectfont{
\setlength{\tabcolsep}{4pt}
\begin{longtable}{@{\extracolsep{\fill}}ccccccccccccc}
\caption{Ionization potentials of the GW100 set\cite{vansettenGW100BenchmarkingG0W02015} obtained from $G_{\text{RSc}}W_0$ and $G_{\text{RSc}}W_{\text{RSc}}$ with HF, BLYP, PBE, B3LYP and PBE0. 
The def2-TZVPP basis set was used.
Systems containing Xe, Rb, I, Ag and Cu were excluded because of the convergence problem.
All values in eV. }\\
\toprule
& & & \multicolumn{5}{c}{$G_{\text{RSc}}W_0$} & \multicolumn{5}{c}{$G_{\text{RSc}}W_{\text{RSc}}$} \\
\cmidrule(l{0.5em}r{0.5em}){4-8} \cmidrule(l{0.5em}r{0.5em}){9-13} 
name & formula & Exp & HF & BLYP & PBE & B3LYP & PBE0 & HF & BLYP & PBE & B3LYP & PBE0 \\
\hline 
\endfirsthead
\caption{Continued}\\
\toprule
& & & \multicolumn{5}{c}{$G_{\text{RSc}}W_0$} & \multicolumn{5}{c}{$G_{\text{RSc}}W_{\text{RSc}}$} \\
\cmidrule(l{0.5em}r{0.5em}){4-8} \cmidrule(l{0.5em}r{0.5em}){9-13} 
name & formula & Exp & HF & BLYP & PBE & B3LYP & PBE0 & HF & BLYP & PBE & B3LYP & PBE0 \\
\hline
\endhead
\bottomrule
\endlastfoot
helium                 & \ce{He}       & 24.59 & 24.59 & 24.24 & 24.29 & 24.35 & 24.39 & 24.58 & 24.65 & 24.69 & 24.64 & 24.66   \\
neon                   & \ce{Ne}       & 21.56 & 21.34 & 21.20 & 21.23 & 21.20 & 21.22 & 21.27 & 21.77 & 21.78 & 21.62 & 21.59   \\
argon                  & \ce{Ar}       & 15.76 & 15.71 & 15.35 & 15.32 & 15.34 & 15.33 & 15.69 & 15.63 & 15.59 & 15.54 & 15.51   \\
krypton                & \ce{Kr}       & 14.00 &       & 13.62 & 13.58 & 13.65 & 13.64 & 13.92 & 13.85 & 13.81 & 13.81 & 13.78   \\
hydrogen               & \ce{H2}       & 15.43 & 16.51 & 16.29 & 16.32 & 16.36 & 16.37 & 16.51 & 16.54 & 16.56 & 16.54 & 16.52   \\
lithium dimer          & \ce{Li2}      & 4.73  & 5.34  & 5.22  & 5.23  & 5.24  & 5.26  & 5.37  & 5.34  & 5.35  & 5.31  & 5.32    \\
sodium dimer           & \ce{Na2}      & 4.89  & 4.99  & 4.98  & 4.98  & 4.98  & 4.98  & 5.00  & 5.03  & 5.04  & 5.00  & 5.00    \\
sodium tetramer        & \ce{Na4}      & 4.27  & 4.30  & 4.18  & 4.19  & 4.19  & 4.21  & 4.31  & 4.22  & 4.23  & 4.21  & 4.23    \\
\end{longtable}
}

\section{Core-Level Binding Energies of the CORE65 Set obtained from Different $GW$ Methods}

\fontsize{9}{11}\selectfont{
\begin{longtable}{@{\extracolsep{\fill}}cccccccc}
\caption{Core-level of the CORE65 set\cite{golzeAccurateAbsoluteRelative2020} obtained from $G_{0}W_{0}$ with HF, BLYP, PBE, B3LYP and PBE0. 
The cc-pVTZ basis set was used.
All values in eV. }\\
\toprule
& & & \multicolumn{5}{c}{$G_{0}W_{0}$} \\
\cmidrule(l{0.5em}r{0.5em}){4-8} 
formula & core level & Ref & HF & BLYP & PBE & B3LYP & PBE0 \\
\hline 
\endfirsthead
\caption{Continued}\\
\toprule
& & & \multicolumn{5}{c}{$G_{0}W_{0}$} \\
\cmidrule(l{0.5em}r{0.5em}){4-8} 
formula & core level & Ref & HF & BLYP & PBE & B3LYP & PBE0 \\
\hline
\endhead
\bottomrule
\endlastfoot
\ce{CH4}       & C1s       & 290.84 & 295.48 &  &  & 285.91 & 286.90 \\
\ce{C2H6}      & C1s       & 290.71 & 295.49 &  &  & 285.85 & 286.89 \\
\ce{C2H4}      & C1s       & 290.82 & 295.90 &  &  & 285.25 & 286.34 \\
\ce{C2H2}      & C1s       & 291.25 & 296.25 &  &  & 285.87 & 287.12 \\
\ce{CO}        & O1s       & 542.10 & 547.89 &  &  & 534.28 &        \\
\ce{CO}        & C1s       & 296.23 & 300.51 &  &  & 290.92 &        \\
\ce{CO2}       & O1s       & 541.32 & 547.39 &  &  & 533.08 & 534.93 \\
\ce{CO2}       & C1s       & 297.70 & 303.15 &  &  & 293.15 & 294.02 \\
\end{longtable}
}

\fontsize{9}{11}\selectfont{
\setlength{\tabcolsep}{4pt}
\begin{longtable}{@{\extracolsep{\fill}}ccccccccccccc}
\caption{Core-level of the CORE65 set\cite{golzeAccurateAbsoluteRelative2020} obtained from ev$GW_{0}$ and ev$GW$ with HF, BLYP, PBE, B3LYP and PBE0. 
The cc-pVTZ basis set was used.
All values in eV. }\\
\toprule
& & & \multicolumn{5}{c}{ev$GW_{0}$} & \multicolumn{5}{c}{ev$GW$} \\
\cmidrule(l{0.5em}r{0.5em}){4-8} \cmidrule(l{0.5em}r{0.5em}){9-13} 
formula & core level & Ref & HF & BLYP & PBE & B3LYP & PBE0 & HF & BLYP & PBE & B3LYP & PBE0 \\
\hline 
\endfirsthead
\caption{Continued}\\
\toprule
& & & \multicolumn{5}{c}{ev$GW_{0}$} & \multicolumn{5}{c}{ev$GW$} \\
\cmidrule(l{0.5em}r{0.5em}){4-8} \cmidrule(l{0.5em}r{0.5em}){9-13} 
formula & core level & Ref & HF & BLYP & PBE & B3LYP & PBE0 & HF & BLYP & PBE & B3LYP & PBE0 \\
\hline
\endhead
\bottomrule
\endlastfoot
\ce{CH4}       & C1s       & 290.84 & 294.05 & 289.58 & 289.72 & 290.43 & 290.89 & 293.74 & 291.53 & 291.72 & 291.83 & 292.11  \\
\ce{C2H6}      & C1s       & 290.71 & 294.00 & 289.39 & 289.36 & 290.33 & 290.83 & 293.65 & 291.39 & 291.63 & 291.73 & 292.03  \\
\ce{C2H4}      & C1s       & 290.82 & 294.33 & 289.17 & 289.68 & 290.64 & 291.16 & 293.95 & 291.87 & 292.02 & 292.08 & 292.34  \\
\ce{C2H2}      & C1s       & 291.25 & 294.69 & 289.70 & 289.72 & 291.14 & 290.82 & 294.35 & 292.07 & 292.32 & 292.44 & 292.76  \\
\ce{CO}        & O1s       & 542.10 & 545.27 & 540.60 & 540.97 &        & 541.49 & 544.72 & 543.11 & 543.57 & 543.33 & 543.87  \\
\ce{CO}        & C1s       & 296.23 & 299.37 & 294.81 & 294.88 &        & 296.00 & 299.08 & 296.56 & 296.72 & 297.04 & 297.26  \\
\ce{CO2}       & O1s       & 541.32 & 544.64 & 539.21 & 540.24 & 540.37 & 539.84 & 543.95 & 541.84 & 542.33 & 542.19 & 542.72  \\
\ce{CO2}       & C1s       & 297.70 & 301.97 & 296.03 & 296.02 & 297.25 & 297.62 & 301.44 & 297.89 & 297.90 & 298.57 & 298.76  \\
\end{longtable}
}

\fontsize{9}{11}\selectfont{
\setlength{\tabcolsep}{4pt}
\begin{longtable}{@{\extracolsep{\fill}}ccccccccccccc}
\caption{Core-level of the CORE65 set\cite{golzeAccurateAbsoluteRelative2020} obtained from $G_{\text{RS}}W_{0}$ and $G_{\text{RS}}W_{\text{RS}}$ with HF, BLYP, PBE, B3LYP and PBE0. 
The cc-pVTZ basis set was used.
All values in eV. }\\
\toprule
& & & \multicolumn{5}{c}{$G_{\text{RS}}W_{0}$} & \multicolumn{5}{c}{$G_{\text{RS}}W_{\text{RS}}$} \\
\cmidrule(l{0.5em}r{0.5em}){4-8} \cmidrule(l{0.5em}r{0.5em}){9-13} 
formula & core level & Ref & HF & BLYP & PBE & B3LYP & PBE0 & HF & BLYP & PBE & B3LYP & PBE0\\
\hline 
\endfirsthead
\caption{Continued}\\
\toprule
& & & \multicolumn{5}{c}{$G_{\text{RS}}W_{0}$} & \multicolumn{5}{c}{$G_{\text{RS}}W_{\text{RS}}$} \\
\cmidrule(l{0.5em}r{0.5em}){4-8} \cmidrule(l{0.5em}r{0.5em}){9-13} 
formula & core level & Ref & HF & BLYP & PBE & B3LYP & PBE0 & HF & BLYP & PBE & B3LYP & PBE0\\
\hline
\endhead
\bottomrule
\endlastfoot
\ce{CH4}       & C1s       & 290.84 & 295.48 & 292.33 & 292.43 & 292.83 & 293.18 & 295.48 & 293.58 & 293.74 & 293.84 & 294.06 \\
\ce{C2H6}      & C1s       & 290.71 & 295.49 & 292.21 & 292.42 & 292.83 & 293.20 & 295.49 & 293.63 & 293.82 & 293.89 & 294.14 \\
\ce{C2H4}      & C1s       & 290.82 & 295.90 & 292.58 & 293.16 & 293.30 & 293.65 & 295.90 & 293.70 & 294.26 & 294.34 & 294.55 \\
\ce{C2H2}      & C1s       & 291.25 & 296.25 & 293.39 & 293.55 & 293.64 & 294.07 & 296.25 & 294.27 & 294.50 & 294.58 & 294.88 \\
\ce{CO}        & O1s       & 542.10 & 547.89 &        & 545.17 &        & 546.34 & 547.89 & 546.63 & 547.06 & 546.81 & 547.25 \\
\ce{CO}        & C1s       & 296.23 & 300.51 &        & 296.57 &        & 297.81 & 300.51 & 298.28 & 298.41 & 298.68 & 298.89 \\
\ce{CO2}       & O1s       & 541.32 & 547.39 & 544.15 &        &        & 544.96 & 547.38 & 545.71 & 546.13 & 545.99 & 546.43 \\
\ce{CO2}       & C1s       & 297.70 & 303.15 & 298.31 &        &        & 299.41 & 303.15 & 299.59 & 299.60 & 300.24 & 300.40 \\
\end{longtable}
}

\fontsize{9}{11}\selectfont{
\setlength{\tabcolsep}{4pt}
\begin{longtable}{@{\extracolsep{\fill}}ccccccccccccc}
\caption{Core-level of the CORE65 set\cite{golzeAccurateAbsoluteRelative2020} obtained from $G_{\text{RSc}}W_0$ and $G_{\text{RSc}}W_{\text{RSc}}$ with HF, BLYP, PBE, B3LYP and PBE0. 
The cc-pVTZ basis set was used.
All values in eV. }\\
\toprule
& & & \multicolumn{5}{c}{$G_{\text{RSc}}W_0$} & \multicolumn{5}{c}{$G_{\text{RSc}}W_{\text{RSc}}$} \\
\cmidrule(l{0.5em}r{0.5em}){4-8} \cmidrule(l{0.5em}r{0.5em}){9-13} 
formula & core level & Ref & HF & BLYP & PBE & B3LYP & PBE0 & HF & BLYP & PBE & B3LYP & PBE0 \\
\hline 
\endfirsthead
\caption{Continued}\\
\toprule
& & & \multicolumn{5}{c}{$G_{\text{RSc}}W_0$} & \multicolumn{5}{c}{$G_{\text{RSc}}W_{\text{RSc}}$} \\
\cmidrule(l{0.5em}r{0.5em}){4-8} \cmidrule(l{0.5em}r{0.5em}){9-13} 
formula & core level & Ref & HF & BLYP & PBE & B3LYP & PBE0 & HF & BLYP & PBE & B3LYP & PBE0 \\
\hline
\endhead
\bottomrule
\endlastfoot
\ce{CH4}       & C1s       & 290.84 & 294.35 & 290.54 & 290.68 & 291.19 & 291.59 & 294.14 & 291.90 & 292.08 & 292.21 & 292.48 \\
\ce{C2H6}      & C1s       & 290.71 & 294.29 & 290.40 & 290.68 & 291.15 & 291.57 & 294.05 & 291.83 & 292.04 & 292.16 & 292.45 \\
\ce{C2H4}      & C1s       & 290.82 & 294.64 & 290.91 & 291.41 & 291.52 & 291.94 & 294.38 & 292.08 & 292.34 & 292.44 & 292.77 \\
\ce{C2H2}      & C1s       & 291.25 & 295.01 & 291.75 & 291.75 & 291.92 & 292.41 & 294.78 & 292.46 & 292.71 & 292.82 & 293.14 \\
\ce{CO}        & O1s       & 542.10 & 545.82 & 542.14 & 542.96 & 543.09 & 543.90 & 545.47 & 543.86 & 544.32 & 544.10 & 544.57 \\
\ce{CO}        & C1s       & 296.23 & 299.56 & 295.56 & 295.64 & 297.10 & 296.57 & 299.37 & 296.87 & 297.01 & 297.30 & 297.54 \\
\ce{CO2}       & O1s       & 541.32 & 545.20 & 541.38 & 541.67 & 542.36 & 542.57 & 544.77 & 542.66 & 543.13 & 543.02 & 543.51 \\
\ce{CO2}       & C1s       & 297.70 & 302.15 & 296.79 & 296.78 & 297.84 & 298.16 & 301.82 & 298.12 & 298.15 & 298.81 & 298.99 \\
\end{longtable}
}

\section{Dipole Moments of Small Molecules Obtained from Different $GW$ Methods}

\fontsize{9}{11}\selectfont{
\begin{longtable}{@{\extracolsep{\fill}}cccccccccccc}
\caption{Dipole Moments of LiH, HF, LiF and CO obtained from DFT and $G_0W_0$ with HF, BLYP, PBE, B3LYP and PBE0. 
Reference values were taken from Ref.\citenum{kaplanQuasiParticleSelfConsistentGW2016}.
The def2-TZVPP basis set was used.
All values in Debye. }\\
\toprule
& & \multicolumn{5}{c}{DFT} & \multicolumn{5}{c}{$G_0W_0$} \\
\cmidrule(l{0.5em}r{0.5em}){3-7} \cmidrule(l{0.5em}r{0.5em}){8-12} 
  & Ref & HF & BLYP & PBE & B3LYP & PBE0 & HF & BLYP & PBE & B3LYP & PBE0 \\
\hline 
\endfirsthead
\caption{Continued}\\
\toprule
& & \multicolumn{5}{c}{DFT} & \multicolumn{5}{c}{$G_0W_0$} \\
\cmidrule(l{0.5em}r{0.5em}){3-7} \cmidrule(l{0.5em}r{0.5em}){8-12} 
  & Ref & HF & BLYP & PBE & B3LYP & PBE0 & HF & BLYP & PBE & B3LYP & PBE0 \\
\hline
\endhead
\bottomrule
\endlastfoot
LiH & 5.88 & 6.02 & 5.56 & 5.60 & 5.71  & 5.76 & 6.10 & 5.39 & 5.49 & 5.54  & 5.78 \\
HF  & 1.82 & 1.95 & 1.80 & 1.80 & 1.85  & 1.85 & 1.63 & 1.78 & 1.81 & 1.69  & 1.78 \\
LiF & 6.28 & 6.49 & 5.95 & 5.97 & 6.11  & 6.16 & 6.38 & 6.08 & 6.04 & 6.21  & 6.28 \\
CO  & 0.11 & 0.30 & 0.17 & 0.21 & 0.07  & 0.08 & 0.18 & 0.51 & 0.47 & 0.00  & 0.22
\end{longtable}
}

\fontsize{9}{11}\selectfont{
\begin{longtable}{@{\extracolsep{\fill}}cccccccccccc}
\caption{Dipole Moments of LiH, HF, LiF and CO obtained from $G_{\text{RS}}W_0$ and $G_{\text{RS}}W_{\text{RS}}$ with HF, BLYP, PBE, B3LYP and PBE0. 
Reference values were taken from Ref.\citenum{kaplanQuasiParticleSelfConsistentGW2016}.
The def2-TZVPP basis set was used.
All values in Debye. }\\
\toprule
& & \multicolumn{5}{c}{$G_{\text{RS}}W_0$} & \multicolumn{5}{c}{$G_{\text{RS}}W_{\text{RS}}$} \\
\cmidrule(l{0.5em}r{0.5em}){3-7} \cmidrule(l{0.5em}r{0.5em}){8-12} 
  & Ref & HF & BLYP & PBE & B3LYP & PBE0 & HF & BLYP & PBE & B3LYP & PBE0 \\
\hline 
\endfirsthead
\caption{Continued}\\
\toprule
& & \multicolumn{5}{c}{$G_{\text{RS}}W_0$} & \multicolumn{5}{c}{$G_{\text{RS}}W_{\text{RS}}$} \\
\cmidrule(l{0.5em}r{0.5em}){3-7} \cmidrule(l{0.5em}r{0.5em}){8-12} 
  & Ref & HF & BLYP & PBE & B3LYP & PBE0 & HF & BLYP & PBE & B3LYP & PBE0 \\
\hline
\endhead
\bottomrule
\endlastfoot
LiH & 5.88 & 6.10 & 5.58 & 5.59 & 5.48 & 5.56 & 6.10 & 5.74 & 5.75 & 5.77 & 5.57 \\
HF  & 1.82 & 1.63 & 1.79 & 1.83 & 1.77 & 1.83 & 1.63 & 1.91 & 1.91 & 1.85 & 1.87 \\
LiF & 6.28 & 6.38 & 6.25 & 6.14 & 6.19 & 6.27 & 6.38 & 6.19 & 6.18 & 6.19 & 6.40 \\
CO  & 0.11 & 0.18 & 0.27 & 0.16 & 0.48 & 0.36 & 0.18 & 0.01 & 0.00 & 0.10 & 0.06
\end{longtable}
}

\fontsize{9}{11}\selectfont{
\begin{longtable}{@{\extracolsep{\fill}}cccccccccccc}
\caption{Dipole Moments of LiH, HF, LiF and CO obtained from $G_{\text{RSc}}W_0$ and $G_{\text{RSc}}W_{\text{RSc}}$ with HF, BLYP, PBE, B3LYP and PBE0. 
Reference values were taken from Ref.\citenum{kaplanQuasiParticleSelfConsistentGW2016}.
The def2-TZVPP basis set was used.
All values in Debye. }\\
\toprule
& & \multicolumn{5}{c}{$G_{\text{RSc}}W_0$} & \multicolumn{5}{c}{$G_{\text{RSc}}W_{\text{RSc}}$} \\
\cmidrule(l{0.5em}r{0.5em}){3-7} \cmidrule(l{0.5em}r{0.5em}){8-12} 
  & Ref & HF & BLYP & PBE & B3LYP & PBE0 & HF & BLYP & PBE & B3LYP & PBE0 \\
\hline 
\endfirsthead
\caption{Continued}\\
\toprule
& & \multicolumn{5}{c}{$G_{\text{RSc}}W_0$} & \multicolumn{5}{c}{$G_{\text{RSc}}W_{\text{RSc}}$} \\
\cmidrule(l{0.5em}r{0.5em}){3-7} \cmidrule(l{0.5em}r{0.5em}){8-12} 
  & Ref & HF & BLYP & PBE & B3LYP & PBE0 & HF & BLYP & PBE & B3LYP & PBE0 \\
\hline
\endhead
\bottomrule
\endlastfoot
LiH & 5.88 & 5.71 & 5.58 & 5.62 & 5.66 & 5.70 & 5.78 & 5.75 & 5.77 & 5.83 & 5.85 \\
HF  & 1.82 & 1.78 & 1.82 & 1.80 & 1.83 & 1.80 & 1.81 & 1.79 & 1.89 & 1.77 & 1.80 \\
LiF & 6.28 & 6.34 & 6.09 & 6.08 & 6.18 & 6.22 & 6.29 & 6.20 & 6.16 & 6.51 & 6.14 \\
CO  & 0.11 & 0.20 & 0.28 & 0.33 & 0.26 & 0.27 & 0.22 & 0.05 & 0.00 & 0.08 & 0.00
\end{longtable}
}

\section{Comparisons of Different RSc Schemes in $G_{\text{RSc}}W_{\text{RSc}}$}

\fontsize{9}{11}\selectfont{
\begin{longtable}{@{\extracolsep{\fill}}ccccc}
\caption{Ionization potentials of the GW100 set\cite{vansettenGW100BenchmarkingG0W02015} obtained from $G_{\text{RSc}}W_{\text{RSc}}$@B3LYP using three schemes to add the correlation correction. 
The def2-TZVPP basis set was used.
Systems containing Xe, Rb, I, Ag and Cu were excluded because of the convergence problem.
All values in eV. }\\
\toprule
& & \multicolumn{3}{c}{$G_{\text{RSc}}W_{\text{RSc}}$@B3LYP} \\
\cmidrule(l{0.5em}r{0.5em}){3-5} 
name & formula & a & b & c \\
\hline 
\endfirsthead
\caption{Continued}\\
\toprule
& & \multicolumn{3}{c}{$G_{\text{RSc}}W_{\text{RSc}}$@B3LYP} \\
\cmidrule(l{0.5em}r{0.5em}){3-5} 
name & formula & a & b & c \\
\hline
\endhead
\bottomrule
\endlastfoot
helium                 & \ce{He}       & 24.64 & 24.62 & 24.62   \\
neon                   & \ce{Ne}       & 21.62 & 21.58 & 21.58   \\
argon                  & \ce{Ar}       & 15.54 & 15.56 & 15.56   \\
krypton                & \ce{Kr}       & 13.81 & 13.80 & 13.78   \\
hydrogen               & \ce{H2}       & 16.54 & 16.52 & 16.52   \\
lithium dimer          & \ce{Li2}      & 5.31  & 5.31  & 5.31    \\
sodium dimer           & \ce{Na2}      & 5.00  & 5.01  & 5.00    \\
sodium tetramer        & \ce{Na4}      & 4.21  & 4.21  & 4.21    \\
\end{longtable}
}

\section{Comparison of Using RSc Orbitals and KS Orbitals in $G_{\text{RSc}}W_{\text{RSc}}$}

\fontsize{9}{11}\selectfont{
\begin{longtable}{@{\extracolsep{\fill}}cccccc}
\caption{Ionization potentials of the GW100 set\cite{vansettenGW100BenchmarkingG0W02015} obtained from $G_{\text{RSc}}W_{\text{RSc}}$ using KS orbitals and using RSc orbitals. 
The def2-TZVPP basis set was used.
Systems containing Xe, Rb, I, Ag and Cu were excluded because of the convergence problem.
All values in eV. }\\
\toprule
& & \multicolumn{2}{c}{$G_{\text{RSc}}W_{\text{RSc}}$@PBE} & \multicolumn{2}{c}{$G_{\text{RSc}}W_{\text{RSc}}$@B3LYP}\\
\cmidrule(l{0.5em}r{0.5em}){3-4} \cmidrule(l{0.5em}r{0.5em}){5-6} 
name & formula & KS orbital & RSc orbital & KS orbital & RSc orbital \\
\hline 
\endfirsthead
\caption{Continued}\\
\toprule
& & \multicolumn{2}{c}{$G_{\text{RSc}}W_{\text{RSc}}$@PBE} & \multicolumn{2}{c}{$G_{\text{RSc}}W_{\text{RSc}}$@B3LYP}\\
\cmidrule(l{0.5em}r{0.5em}){3-4} \cmidrule(l{0.5em}r{0.5em}){5-6} 
name & formula & KS orbital & RSc orbital & KS orbital & RSc orbital \\
\hline
\endhead
\bottomrule
\endlastfoot
helium                 & \ce{He}       & 24.69 & 24.66 & 24.64 & 24.62  \\
neon                   & \ce{Ne}       & 21.78 & 21.74 & 21.62 & 21.58  \\
argon                  & \ce{Ar}       & 15.59 & 15.56 & 15.54 & 15.56  \\
krypton                & \ce{Kr}       & 13.81 & 13.81 & 13.81 & 13.80  \\
hydrogen               & \ce{H2}       & 16.56 & 16.54 & 16.54 & 16.52  \\
lithium dimer          & \ce{Li2}      & 5.35  & 5.35  & 5.31  & 5.31   \\
sodium dimer           & \ce{Na2}      & 5.04  & 5.04  & 5.00  & 5.00   \\
sodium tetramer        & \ce{Na4}      & 4.23  & 4.23  & 4.21  & 4.21   \\
\end{longtable}
}

\bibliography{ref}